\documentclass[sigconf,screen]{acmart}

\acmConference[ISSTA 2025]{The ACM SIGSOFT International Symposium on Software Testing and Analysis }{June 2025}{Trondheim, Norway}

\acmBooktitle{The ACM SIGSOFT International Symposium on Software Testing and Analysis (ISSTA '25), June 23--27, 2025, Trondheim, Norway}

\def\BibTeX{{\rm B\kern-.05em{\sc i\kern-.025em b}\kern-.08emT\kern-.1667em\lower.7ex\hbox{E}\kern-.125emX}}

\usepackage{amsfonts}
\usepackage{graphicx}
\usepackage{array}
\usepackage{url}
\usepackage{fancybox}
\usepackage{multirow}
\usepackage{color}
\usepackage{colortbl}
\usepackage{balance}
\usepackage{fancyhdr}
\usepackage{diagbox}
\usepackage{bbding}
\usepackage{placeins}
\usepackage{tabularx}
\usepackage{enumitem}
\usepackage{xcolor,lipsum}
\usepackage{textcomp}
\usepackage{algorithmic}
\usepackage{listings}
\usepackage{color}
\usepackage{commath}
\usepackage{comment}
\usepackage{bm}
\usepackage{amsmath,amsfonts}
\usepackage{algorithmic}
\usepackage{tcolorbox}
\usepackage{graphicx}
\usepackage{listings}
\usepackage{makecell}
\usepackage{adjustbox}
\usepackage{multirow}
\usepackage{graphicx}
\usepackage{svg}
\usepackage{xspace}
\usepackage{dsfont}
\usepackage{lipsum}
\usepackage{centernot}
\usepackage{diagbox}
\usepackage{tabularx}

\newcolumntype{P}[1]{>{\centering\arraybackslash}p{#1}}
\usepackage[ruled,vlined,linesnumbered]{algorithm2e}

\SetCommentSty{mycommfont}
\usepackage{pifont}
\usepackage{enumitem}
\usepackage{caption}

\usepackage{geometry}
\usepackage{pgf-pie}
\usepackage{multicol}
\usepackage{listings}
\usepackage{ragged2e}

\definecolor{light-gray}{rgb}{.906,  .902,  .902}

\newcommand{\toolname}{\textsc{Code2API}\xspace}

\begin{document}
\title[]{Code2API: A Tool for Generating Reusable APIs from Stack Overflow Code Snippets}


\author{Yubo Mai}
\affiliation{%
\institution{Zhejiang University}
\country{China}
}  
\email{12021077@zju.edu.cn}

\author{Zhipeng Gao}
\affiliation{%
\institution{Zhejiang University}
\country{China}
}
\email{zhipeng.gao@zju.edu.cn}

\author{Xing Hu}
\affiliation{%
\institution{Zhejiang University}
\country{China}
}
\email{xinghu@zju.edu.cn}

\author{Lingfeng Bao}
\affiliation{%
\institution{Zhejiang University}
\country{China}
}
\email{lingfengbao@zju.edu.cn}

\author{Jingyuan Chen}
\affiliation{%
\institution{Zhejiang University}
\country{China}
}
\email{jingyuanchen@zju.edu.cn}

\author{Jianling Sun}
\affiliation{%
\institution{Zhejiang University}
\country{China}
}
\email{sunjl@zju.edu.cn}

\begin{abstract}
Nowadays, developers often turn to Stack Overflow for solutions to daily problems, 
however, these code snippets are partial code that cannot be tested and verified properly. 
One way to test these code snippets is to transform them into APIs (\textbf{A}pplication \textbf{P}rogram \textbf{I}nterface) that developers can directly invoked and executed. 
However, it is often costly and error-prone for developers to manually perform this transformation (referred to as AIPzation task) due to different actions to be taken (e.g., summarizing proper method names, inferring input parameters list and return statements). 
To help developers quickly reuse code snippets in Stack Overflow, in this paper, we propose \toolname, a Google Chrome extension that uses Large Language Models (LLMs) to automatically perform APIzation of code snippets on Stack Overflow. 
\toolname guides LLMs through well-designed prompts to generate reusalbe APIs, using Chain-of-Thought reasoning and few-shot in-context learning to help LLMs understand and solve the APIzation task in a developer-like manner. 
The evaluation results show that \toolname significantly outperforms the rule-based approach by a large margin. 
The full paper of this tool has been published in FSE’24 as a research paper~\cite{mai2024human}.

\noindent Demo video: \url{https://youtu.be/RI-ZpBnNNwQ}. \\
\noindent Demo website: \url{https://doi.org/10.6084/m9.figshare.24426961.v1}. \\
\noindent Replication package: \url{https://github.com/qq804020866/Code2API}.

\end{abstract}

\maketitle

\section{INTRODUCTION}
\label{sec:intro}
Stack Overflow, a leading Software Q\&A site, is a valuable resource for developers seeking answers to technical their queries~\cite{gao2020generating, gao2023know, gao2020technical}. 
Code snippets shared in Stack Overflow provide practical solutions to daily programming problems. 
However, the Stack Overflow code snippets are primarily used for illustrative purposes and are not able to be reused directly. 
To use these code snippets more effectively, developers have to convert them into more user-friendly formats, such as APIs. 
Creating a reusable API from Stack Overflow code snippets, namely APIzation, is a non-trivial task. 
For example, a skilled developer has to perform the necessary actions, such as summarizing proper method names, resolving compilation issues, inferring input parameters and return statements, etc. 
The modification process is time-consuming and error-prone. 

To automate APIzation, Terragni et al.~\cite{terragni2021apization} developed a tool called APIzator, which converts Java code snippets into compilable APIs. 
However, according to our investigation, 185 (92.5\%) APIs generated by APIzator are not applicable due to their imprecise and misleading method names. 
Moreover, the rules designed for APIzator are too complex and specific, making it hardly applicable to other programming languages. 

Inspired by the recent success of large language models (LLMs) in code understanding~\cite{chen2021evaluating, xue2024selfpico, mai2024towards, huang2024let}, we propose \toolname, a Google Chrome extension to automatically generate reusable and applicable APIs for Stack Overflow code snippets. 
\textbf{\toolname uses prompt engineering, chain-of-thought reasoning, and few-shot in-context learning to guide the LLMs to accomplish the APIzation task in a manner similar to a skilled developer}. 
Specifically, the chain-of-thought reasoning includes the key steps that skilled developers take to accomplish the APIzation task. 
The examples in few-shot learning show the LLMs the expected input and output of the APIzation task, making it easier for the model to quickly understand the task. 
In addition, we have also designed role designation and format constraints to assist LLMs in completing the APIzation task and control the output format for subsequent processing. 

We empirically evaluated the performance of \toolname from three aspects: the accuracy of \toolname generated input parameters, return statements, and equivalent method implementations;  
the quality of method names and the overall APIs generated by \toolname; and the generalization ability of \toolname. 
Extensive evaluations indicate that: (1) For 66.0\% and 65.0\% generated APIs, APIzator and human expert extracted identical parameters and return statements respectively. 
For 43.5\% APIs, \toolname generated equivalent method implementations with humans, which are 11\%, 9.5\%, and 7.5\% higher than APIzator, respectively. 
(2) In terms of method names, 
74.3\% of the Java method names generated by \toolname are descriptive and meaningful (e.g., scored as 4), which is far superior than APIzator (10\%) and even better than human developers (60\%). 
Regarding the overall API quality evaluation, 50.5\% developers choose the APIs generated by our approach as the best ones, while 48\% developers choose human-written APIs as their first choice, and only 1.5\% developers consider APIzator's results as the best. 
This suggests that \toolname has achieved a level comparable or even superior to human developers in the task of Java APIzation. 
(3) Finally, we slightly adjusted the prompt for Java APIzation to enable \toolname to complete the Python APIzation task. 
The results show that \toolname has an accuracy of 69.0\%, 80.0\%, and 57.0\% for parameters, return statements, and equivalent method implementations on the Python dataset, respectively. 
The consistently promising performance proves that \toolname can easily extend to other programming languages without losing performance.





\section{APPROACH}
\label{sec:approach}
\begin{figure}[h]
	\centering
	\includegraphics[width = 0.95\linewidth]{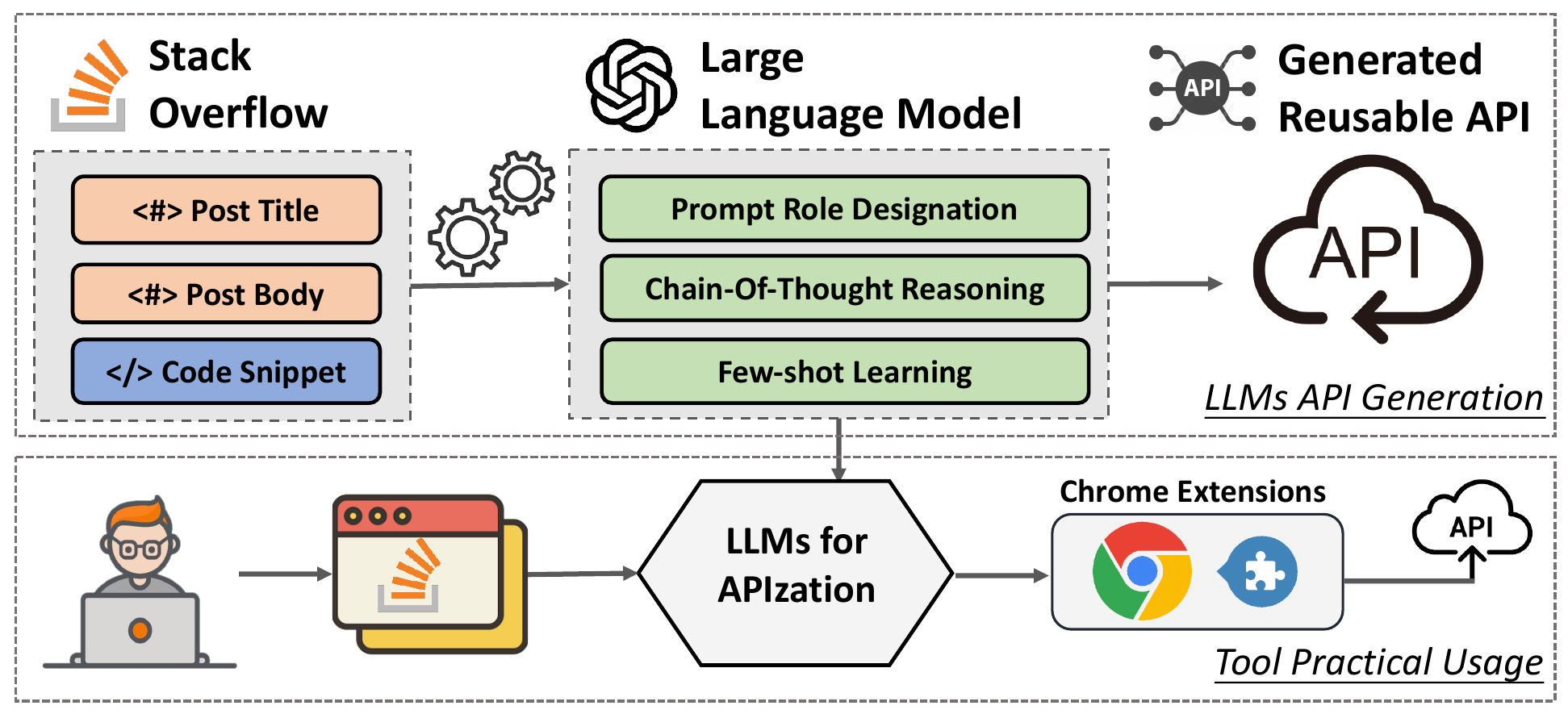}
	\caption{The Workflow of Our Approach.}
	\label{fig:workflow}
\end{figure}

In this section, we introduce the method details of \toolname. The framework of this method is depicted in Fig.~\ref{fig:workflow}. 
As shown in Fig.~\ref{fig:workflow}, we first extract the question title, question body, and answer body from a given Stack Overflow post. 
We then use the above information to fill our designed prompt for APIzation task. 
Since LLMs are not specifically designed for the APIzation task, we use prompt role designation, chain-of-thought reasoning, and few-shot in-context learning strategy to guide LLMs to produce the desired output. 
Our model is extremely lightweight and flexible, as it doesn't require any manually designed rules or extensive training data.

\subsection{\label{2.1}Data Preparation}
Our task focuses on Stack Overflow code snippets, so we start by downloading the official Stack Overflow data dump from StackExchange. 
This data dump includes timestamped information about the posts, each of which contains a brief question title, a detailed question post, corresponding answers, and multiple tags. 
The code snippets in answer posts offer solutions to specific technical problems.  
To help LLMs better understand the code context and target problems, 
for a given Stack Overflow answer code snippet (enclosed by <code> tags), we retrieve its associated question title, question post and answer post, providing enough context for LLMs to understand the problem purpose and code semantics.

\subsection{Prompt Engineering}
\toolname primarily uses prompt engineering, which prompts LLMs to generate expected reusable APIs. 
The LLMs have been trained on ultra-large-scale data, which learn comprehensive language and code knowledge and can be effectively guided through reasonable prompts. 
By simply describing natural language prompts, LLMs can perform specific tasks without additional training or hard coding. 
Prompt engineering has been successfully used in various software engineering tasks, such as code generation~\cite{dong2023self, dai2024mpcoder}, program repair~\cite{paulenhancing, dai2025less}, and code comprehension~\cite{yuan2023evaluating, xing2025prompt}. 
In this work, we use prompt role designation, chain-of-thought reasoning, and few-shot in-context learning to utilize LLMs’ knowledge for automated API construction for Stack Overflow code snippets.

\subsubsection{Prompt Role Designation.}
In prompt engineering, role designation is a technique where LLMs are assigned a role to solve a specific task. Giving a role to LLM provides it with the problem context that aids its understanding of the task context, and results in more accurate and relevant responses. 
In this study, as we aim to convert code snippets (Java in this case) into reusable APIs, we assign the role of LLMs to act as a skilled Java developer. 
After assigning the role to LLMs, we clearly inform the LLMs about our task as follows: 
 "\textit{Give you a context including a question title, a question post and an answer post, your task is to transform the java code snippet in the answer post into a java method based on the context}".
The prompt role designation clearly indicates the input and output of the task, drawing on the programming knowledge of LLMs to complete this APIzation task.

\subsubsection{Chain-of-Thought Reasoning.}
Chain-of-thought reasoning is a crucial method for prompt engineering. 
It allows LLMs to break down a complicated task into several simpler steps, producing a sequence of intermediate results that ultimately lead to a coherent and logical outcome. 
The process of converting a code snippet into a reusable API, which is a complex task, necessitates a series of intermediate steps to generate the final API. 
To implement this approach, two developers were requested to independently document their chain-of-thought process for converting 15 Stack Overflow Java code snippets, in line with previous research~\cite{feng2023prompting,wei2022chain}. 
The first author then discussed with the two developers to determine the most effective reasoning for each post, summarizing a clear, sequential thought process that leads to generating the final reusable APIs. 
Specifically, we established an 8-step thought process of developers as the chain-of-thought reasoning for LLMs. 
This allows LLMs to convert code snippets using the same thought process as experienced developers. 
Specifically, step 1 in the chain-of-thought is "\textit{Recover import statements based on the code snippet. If necessary, it can be none}". 
This step is used to solve compilation errors due to missing type declarations. 
Step 2 is "\textit{Define a public class Chatgpt that will be used to wrap the method}", which creates a class named Chatgpt to facilitate subsequent processing. 
Step 3 is "\textit{Create 'public static' modifier for the method}",  providing the method with a default modifier.
The 4th to 7th steps are, "\textit{Create the method name based on the context or the code snippet itself}", "\textit{Infer parameter list based on the code snippet. If necessary, it can be none}", "\textit{Infer return statements based on the code snippet. If necessary, it can be none}", "\textit{Infer throws statements based on the code snippet. If necessary, it can be none}" respectively. 
These steps guide LLMs to sequentially solve sub-tasks such as creating the method name, inferring the parameter list, return statements and throws statements. 
Step 8 is "\textit{Output the complete code based on above results}", which is used to generate the entire API.
Overall, this sequential thought process guides LLMs to produce reusable APIs step by step in a manner similar to skilled developers.

\subsubsection{Few-Shot Learning.}
The growing capabilities of LLMs have led to the widespread use of in-context learning in the form of zero-shot learning and few-shot learning. 
Since LLMs are not specifically trained using Stack Overflow code snippets, we employ the few-shot learning approach as our in-context learning paradigm. 
This method involves enhancing the context with a number of desired input and output examples, aiding the model in extracting the necessary knowledge and abstractions to accomplish the task. 
In particular, for each step within the Chain-of-thought, we demonstrate LLMs with each step's input and output in terms of a given example. 
To choose the most representative examples for Java APIzation, the first author examined 100 Stack Overflow posts and selected those that covered most of the steps in the chain-of-thought reasoning process. 
In the end, we selected five examples that best represent our needs, and through further comparison of the performance of different n-shot learning, we ultimately chose 1-shot learning to construct our final prompt.

\subsubsection{Prompt construction.}
The final prompt we've constructed is made up of six components: prompt role designation, chain-of-thought reasoning, example input, example output, test input and format constraints. Each component has a unique function as described below:
\begin{itemize} 
\item\textbf{Prompt Role Designation:} 
It provides a detailed summary of the APIzation task for Stack Overflow code snippets, establishing the context for the following steps.
\item\textbf{Chain-of-Thought Reasoning:} It guides the LLMs in a step-by-step manner, allowing the LLM to tackle the APIzation task using the same thought process as developers.
\item\textbf{Example Input and Example Output:} These examples further clarify the task's requirements and demonstrate the expected output format, helping the model to comprehend the expectations more clearly. 
\item\textbf{Test Input:} This component represents the problem that our model is currently solving, acting as a real-world evaluation scenario. 
\item\textbf{Format Constraints:} It specify the exact input format for the code snippet APIzation task and the desired output format of the LLMs for subsequent data processing. 
\end{itemize}

These six components are combined to form our final prompt, a full version of which is included in our replication package: 
\url{https://github.com/qq804020866/Code2API}.
Once the constructed prompt is fed into the LLMs, the LLMs will produce the APIzation result. 
We employ regular expressions to post-process the LLMs' output and save the generated APIs in local files for further assessment.  


\section{Tool Availability}
\label{sec:application}
We have implemented \toolname as a Google Chrome extension, which can be downloaded at \url{https://doi.org/10.6084/m9.figshare.24426961.v1}. 
We demonstrated the process of installing \toolname Google Chrome extension and how to use our tool in practice in our tool demo video \url{https://youtu.be/RI-ZpBnNNwQ}. 
\toolname Google Chrome extension consists of three components: a frontend part, a middleware part, and a backend part. 
The frontend model runs on the user's browser, which is responsible for fetching the user-interested code snippets from Stack Overflow webpages and displaying the generated results. 
The middleware part combines the necessary information from the frontend data and automatically constructs prompt through our prompt engineering process (detailed in Section~\ref{sec:approach}). 
After the prompt is constructed, our model sends the prompt to the backend model (LLMs) to generate reusable APIs. 
Finally, the frontend alters the page to present the generated APIs to developers. 
\textbf{Any users with a Chrome browser can install and use our tool.} 

\begin{figure} 
	\centering
	\includegraphics[width = 0.95\linewidth]{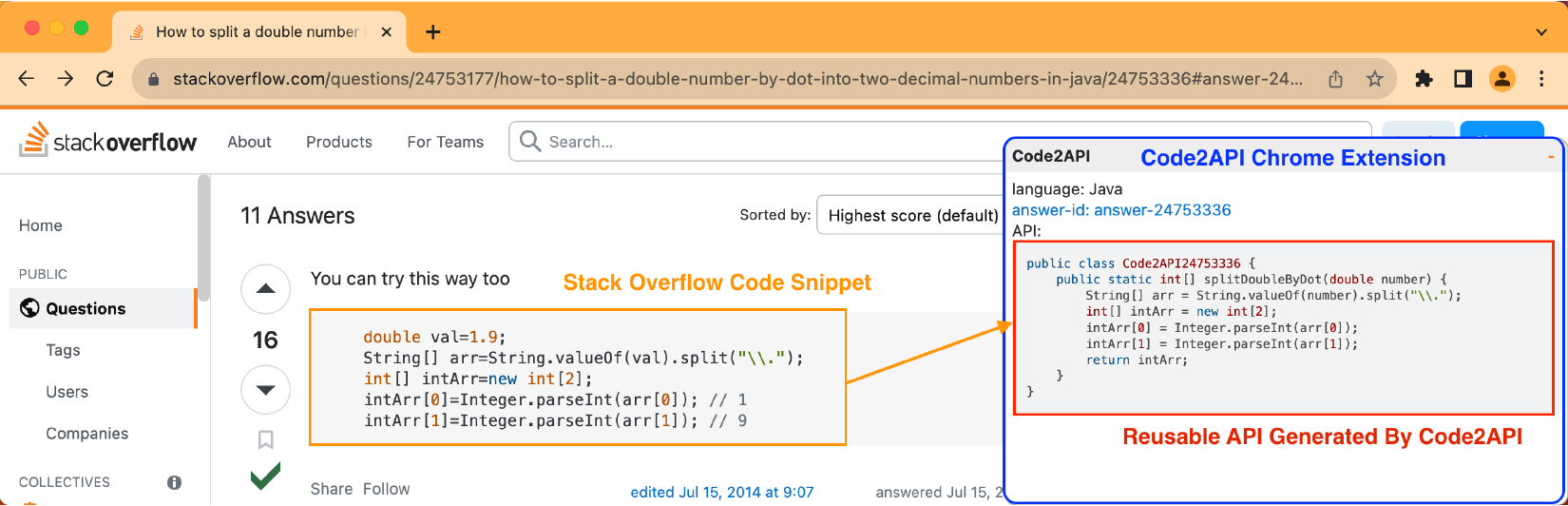}
	\caption{The Overview of {\sc Code2API} Chrome Extension}
	\label{fig:extension}
\end{figure}

\noindent\textbf{Frontend Model:} 
The frontend has two main features: fetching code snippets and necessary code context information from the webpage and displays the generated results. 
(1) Fetching code snippet: When loading stack overflow pages (Java and Python for this study), the frontend will fetch the user-interested code snippets and its associated code context information. 
When a user clicks on a code snippet in an answer post, \toolname will collect all the necessary code context as described in Section~\ref{2.1}, and sends them to the middleware model for parsing. 
(2) Displaying results: After the backend model finishes the APIzation, the frontend interface will display the generated reusable APIs and format the APIs to the clipboard for users to take away. The demonstration of the frontend interface is shown in Figure~\ref{fig:extension}.

\noindent\textbf{Middleware Model:} The middleware model takes the necessary information (e.g., code snippet, question title, question body, answer body) from the frontend model, and constructs prompt following our prompt engineering process (detailed in Section~\ref{sec:approach}). 
Then the middleware model sends the constructed prompt to the backend model for inferecing. 

\noindent\textbf{Backend Model:} 
The backend model is a LLM and is responsible for analyzing the code snippets and generating the reusable APIs for the Stack Overflow code snippets. 
In this study, We selected the cutting-edge GPT-3.5-turbo model\footnote{\url{https://platform.openai.com/docs/model-index-for-researchers}} as our LLM, which is recognized as one of the top instruction-tuned LLMs~\cite{ouyang2022training}. 
This model has demonstrated superior performance in areas like text summarization~\cite{yang2023exploring} and machine translation~\cite{hendy2023good}. 
To maintain the distinctiveness of our experimental outcomes, we adjusted the temperature parameter to zero throughout all experiments, ensuring consistent evaluation output. 
All the prompts we created adhere to the maximum input tokens limit of GPT-3.5-turbo, indicating that the GPT-3.5-turbo model is adequately equipped to solve this APIzation task. 

\noindent\textbf{Model Efficiency:} As reported by Terragni et al.~\cite{terragni2021apization}, developers need 4min and 22s on average to perform a single APIzation. 
In our evaluation experiments, we found that the average time for \toolname to complete one APIzation is 10 seconds, which is significantly less than the time used by developers. 
This implies that \toolname can greatly improve the developer's efficiency in reusing code snippets, and allows developers to incorporate reusable APIs in their daily development just-in-time.

\section{Evaluation}
\label{sec:eval}


\subsection{Parameter Lists, Return Statements, and Equivalent Methods}
We compared the performance of \toolname with the state-of-the-art model APIzator using the same evaluation set by Terragni et al.~\cite{terragni2021apization} (containing 200 APIs written by developers as ground truth). 
The evaluation results showed that the accuracy of generated input parameters, return statements, and equivalent method implementations of \toolname were 66.0\%, 65.0\%, and 43.5\%, respectively, which were 11\%, 9.5\%, and 7.5\% higher than those of APIzator. 
This is because \toolname is not limited to complicated designed rules and can generate APIs for a wide range of code snippets, thus improving accuracy.

\subsection{Method Names and the Overall API}
We recruited 18 volunteers to conduct a user study to evaluate the method names and overall API quality generated by \toolname, APIzator and human developers. 
We divided the entire dataset (200 triplets) into 6 groups, and each group was evaluated by 3 volunteers. 
To evaluate method names, we used scores from 1-4 to represent the descriptiveness between method names and its method bodies, with higher scores indicating higher relevance (a score of 4 means the method name matches the method implementation perfectly). 
To evaluate overall API quality, each evaluator chose the best API among the three candidates, and the best API is determined by majority voting. 
It is worth mentioning that the evaluator did not know which API was generated by which method. 

The evaluation results of method names are shown in Table~\ref{tab:RQ2_R}. 
It can be seen that 74.3\% of the APIs generated by Code2API scored 4 points, far superior than the results of APIzator (10.0\%), and even surpassed the human level (60.0\%). 
The average scores of the three methods also showed the same trend. 
\begin{table} 
\footnotesize
\caption{Method Name Evaluation Results}
\label{tab:RQ2_R}
\centering
\begin{tabular}{llllll}
\toprule
\textbf{MNE SCORE} & 1 & 2 & 3 & 4 & Avg \\
\hline
Human & 25 (4.2\%) & 77 (12.8\%) & 138 (23.0\%) & 360 (60.0\%) & 3.39 \\
APIzator & 136 (22.7\%) & 248 (41.3\%) & 156 (26.0\%) & 60 (10.0\%) & 2.23 \\
\toolname & 14 (2.3\%) & 29 (4.8\%) & 111 (18.5\%) & 446 (74.3\%) & 3.65 \\
\bottomrule
\end{tabular}
\end{table}

The best API evaluation results show that among the 200 best APIs, 101 (50.5\%) were generated by Code2API, 96 (48.0\%) were generated by human developers, and 3 (1.5\%) were generated by APIzator. 
This evaluation demonstrates that \toolname has achieved a level comparable or even superior performance to human developers in the APIzation task. 

We attribute the promising performance of \toolname to: (1) the remarkable performance of natural language understanding and logical reasoning abilities of LLMs; (2) we construct reasonable prompts for this task with chain-of-thought reasoning and in-context learning. (3) we provide sufficient context (e.g., question title, question body and answer body) to LLMs for inference.

\subsection{Generalization Study}
To demonstrate that \toolname generalization ability, we first manually constructed a ground-truth dataset containing 100 Python APIs following the data collection process of APIzator and manually wrote reusable APIs. 
We then slightly modified \toolname by changing the few-shot examples and chain-of-thought reasoning from Java to Python. 
The results show that the accuracy of parameter lists, return statements, and equivalent method implementation of \toolname in Python APIzation were 69.0\%, 80.0\%, and 57.0\%, respectively, all of which were better than the corresponding measures in Java APIzation. 
A possible reason is that Python variables do not need to explicitly declare types, so it is easier to infer parameter lists and return statements. 
The consistently promising performance of \toolname on the Python dataset proves that our framework can be easily extended to other programming languages without sacrificing performance. Considering the effectiveness of \toolname for this APIzation task, we use our tool to create two large-scale dataset for Stack Overflow code snippets, including 6023 Java reusable APIs and 5000 Python reusable APIs.



\section{Conclusion and future work}
\label{sec:con}
This paper proposed a Google Chrome extension, \toolname, to help developers quickly convert code snippets to reusable APIs in Stack Overflow. 
The backend of \toolname utilizes chain-of-thought reasoning and few-shot learning to guide LLMs for inference in a manner similar to developers' thought processes. 
The frontend allows developers to select their interested code snippets and displays the final reusable APIs. 
The evaluation results show \toolname achieves a comparable or even superior performance than human developers on this task. 
Our tool is highly efficient and can easily generalize to other programming languages without sacrificing performance. 
We will make \toolname support other programming languages in future work.

\section*{ACKNOWLEDGMENT}
\label{sec:ack}

This research is partially supported by the National Science Foundation of China (No. 62202341), the Shanghai Sailing Program (23YF1446900), and the CCF-Tencent Rhino-Bird Open Research Fund (No. RAGR20240). 
This research is partially supported by the Ningbo Natural Science Foundation (No. 2023J292). 
The authors would like to thank the reviewers for their insightful and constructive feedback.


\balance
\bibliographystyle{ACM-Reference-Format}
\bibliography{samples}

\end{document}